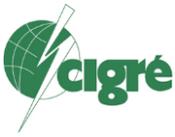
21, rue d'Artois, F-75008 PARIS
http : //www.cigre.org

# Electric Vehicle Battery Swapping Station

M. MAHOOR, Z. S. HOSSEINI, A. KHODAEI  
University of Denver  
USA

D. KUSHNER  
ComEd  
USA

**SUMMARY**

Providing adequate charging infrastructure plays a momentous role in rapid proliferation of Electric Vehicles (EVs). Easy access to such infrastructure would remove various obstacles regarding limited EV mobility range. A Battery Swapping Station (BSS) is an effective approach in supplying power to the EVs, while mitigating long waiting times in a Battery Charging Station (BCS). In contrast with the BCS, the BSS charges the batteries in advance and prepares them to be swapped in a considerably short time. Considering that these stations can serve as an intermediate entity between the EV owners and the power system, they can potentially provide unique benefits to the power system. This paper investigates the advantages of building the BSS from various perspectives. Accordingly, a model for the scheduling of battery charging from the station owner perspective is proposed. An illustrative example is provided to show how the proposed model would help BSS owners in managing their assets through scheduling battery charging time.

Mohsen.Mahoor@du.edu

# 1. INTRODUCTION

The development of electric vehicles (EVs) is widely favoured by a larger and growing segment of car owners, manufacturers, governments, municipalities and investors, mainly due to the potential to reduce reliance on fossil-fuel based resources, and accordingly, reduce consequent environmental impacts [1]. It is anticipated that EVs will take 25% of the automotive market by 2020. Growing penetration of EVs can potentially reduce emission, save fuel cost for EV owners, and reduce the consumption of gasoline. It can also increase utilization of renewable energy such as wind and solar resources, as the EV's battery has the storage ability which can be potentially employed as a flexible source for intermittent energy resources [2].

Battery charging plays a pivotal role in the adaptability of EVs. The current charging scheme is mostly based on plugging the EV into an outlet, either individual outlets or in a Battery Charging Station (BCS), and leave the car for hours to be fully charged. This method takes much longer than fueling a gasoline-powered vehicle and presents a shortcoming and a barrier to EV adoption. On the other hand, the mobility range of the current EVs is in the order of couple of hundred miles. Although this range is sufficient for daily travels, a large number of EV owners consider this as an important factor in their driving needs [3]. Another obstacle, from a BCS owner perspective, is the cost of building charging facilities and the required real estate. Due to the fact that the EVs need enough space to be parked for several hours and be charged, deploying the BCS is costly, especially in densely populated urban areas.

One state-of-the-art solution for overcoming these obstacles is to swap the empty batteries by fully-charged ones [3]. The idea is that EV owners can easily pull over in a swapping station where an empty battery is automatically switched with a fully-charged one. Different from charging EVs in the plug-in method, battery swapping approach just takes a few minutes to replace a battery [4]. By determining the optimal locations of BSS, drivers not only could charge their EVs as fast and easy as refuelling in a gas station, but also could extend their travel distances.

To fully realize the potential of the BSS, the EVs' batteries should be easily replaceable and accessible [3]. One important requirement would be to establish consistent standards for the batteries of various EVs. [1]. In this respect, the best model of battery possession is the company-owned battery model in which EV owners lease the batteries while they are owned by the company [4], [5]. The most prominent feature of this approach, in addition to the reduced charging time, is that the price of the EVs dramatically drops, as the cost of the battery is deducted from the total vehicle cost.

This paper investigates various aspects of deploying the BSS solution from the perspective of the EV owner, station owner and the power system operator. A model from the station owner perspective is then proposed with the objective of scheduling battery charging, while taking into account the prevailing constraints. The remainder of this paper is organized as follows: Section 2 studies this issue from the various perspectives. Section 3 proposes the BSS battery charging model. An illustrative example on a test BSS is provided in Section 4. Section 5 concludes the paper.

## 2. BSS BENEFITS

As depicted in Figure 1, the BSS approach necessitates mutual interactions with all players involved, including but not limited to the EV owner, the station owner, and the power system. An EV owner would like to swap his/her empty battery with a fully-charged one within a few minutes, while the station owner considers the electricity price to charge the empty batteries and minimize the associated cost. The inherent interdependency of the swapping station and the power system is significant. In what follows, the issue of BSS and its associated benefits are studied from various viewpoints.

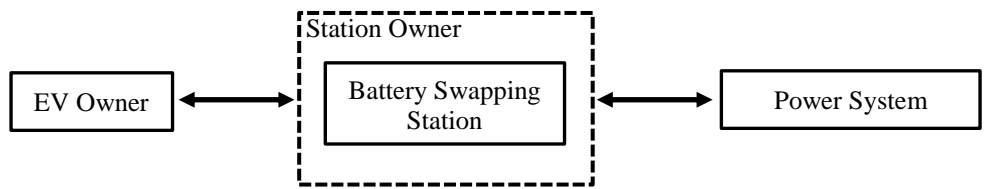

Figure 1. Interactions of BSS with EV owners, station owner and power system.

**POWER SYSTEM PERSPECTIVE:**
As the behaviour of EV owner is stochastic and unpredictable, there is no chance to apply a controlled strategy for charging and discharging of the batteries when an EV is under the plug-in mode. The uncontrolled nature of charging facilities in a plug-in mode could have significant impacts on the power system, such as increasing peak load and network congestion. The BSS approach, however, offers a controlled charging strategy in terms of scheduling battery charging time. It means that the BSS is able to postpone the charging of batteries to the night time or off-peak hours, and accordingly, the aforementioned problem will be solved [5], [6]. From the power system perspective, the BSS can be treated as a large flexible load. By controlling the charging and discharging time of the batteries, the potential peak demand or overloading, caused by increasing penetration of EVs, can be flattened. It can be achieved by determining an intelligent charging schedule without the need of upgrading the current grid infrastructure [6], [7].

**EV OWNER PERSPECTIVE:**
The first and foremost objective of employing EV is to provide the required mobility service anticipated by the EV owner. EVs need to attain their electric energy from the grid and store it in their batteries and further utilize while on the move [6-8]. From the EV owner's viewpoint, the BSS has various key advantages: 1) reducing the sticker price of EVs, as the station owns the batteries, 2) speeding up the battery charging, which would become as fast as refuelling a gasoline-powered vehicle, 3) allowing longer trip distance for the EV owners by accessing the fast battery swapping in the BSS, 4) relieving the concern of battery lifetime as the BSS operator runs healthy and advanced control strategy for battery charging [9] to avoid sequential damages, and 5) decreasing the cost of upgrading household infrastructure to high power chargers.

**STATION OWNER PERSPECTIVE:**
In line with the rapid development of battery technology, the price of battery and its associated facilities will significantly drop, which causes the BSS to be more practical. The BSS owner not only should know the initial number of batteries to be purchased, but also should charge the batteries in a time-scheduled manner on the basis of electricity prices [3]. Moreover, the storage capability of batteries provides a great opportunity for the BSS owner to offer grid services. Therefore, deploying BSS offers various benefits for its owner, some of which are 1) minimizing its electricity cost by scheduling the battery charging process, 2) maximizing its profit by participating in electricity markets and also providing ancillary services, such as demand response and spinning reserve, 3) reducing the cost of real estate, as there is no need to access large parking spaces, and 4) offering convenience for charging the batteries due to the availability of consistent battery standards.

## 3. BSS SCHEDULING MODEL

A BSS charging scheduling model is investigated in this section. The proposed model aims at scheduling the battery charging with respect to the availability of battery chargers, and hourly demand for swapping the batteries. As shown in Figure 2, four different states are considered for each battery: empty ($x^E$), charging ($x^C$), fully-charged ($x^F$), and out-of-station ($y$). Each of these states is modelled using a binary variable. If a binary variable in 1 it means that the battery is in that state. As a battery reaches the fully-charged state, it will be ready to be swapped based on the demand. In addition, once an empty (or not-fully-charged) battery is delivered to the station, it will be queued, and further moved forward to be charged based on the availability of charging facilities and demand.

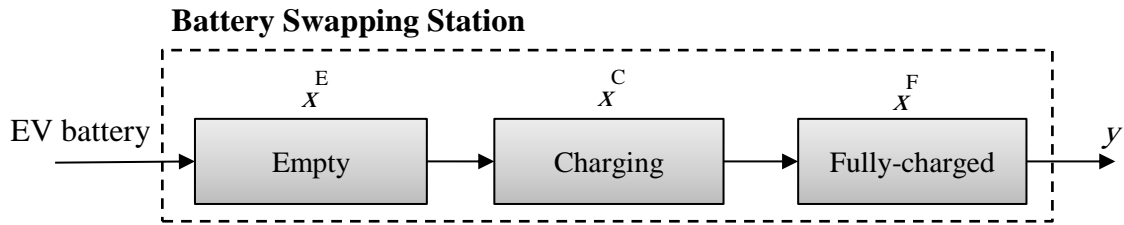

Figure 2. Battery states in BSS.

At each hour, every individual battery transits among these four states, as defined in (1):

$$(x_{bt}^E + x_{bt}^C + x_{bt}^F) + y_{bt} = 1 \qquad \forall t, \forall b \qquad (1)$$

As each battery enters the station, its state is toggled in a sequential order. When an empty battery is delivered from an EV owner, it will be put under the empty state, denoting $x^E = 1$. Once there is an empty slot to move this battery forward, the charging will start, $x^E$ will become zero, and accordingly $x^C$ will be one. When the battery is fully-charged, it will be moved to the next state, i.e., fully-charged, thus $x^F = 1$ and $x^C = 0$. A battery in the fully-charged state is ready to be swapped. By handing over the fully-charged battery to the EV owner, the battery is going out of the station, and consequently $y$ will become one, while other states ($x^E$, $x^C$ and $x^F$) will be zero.

Equation (2) demonstrates that the total number of batteries, associated with this BSS, is a constant, represented by $N^S$. In other words, the station owner possesses this number of batteries to provide the swapping service to EV owners.

$$\sum_b (x^E_{bt} + x^C_{bt} + x^F_{bt} + y_{bt}) = N^S \qquad \forall t \qquad (2)$$

The maximum number of batteries that can be charged simultaneously is limited by the number of charger in BSS. Denoting the number of battery chargers by $N^C$, the associated limit can be modelled is as (3):

$$\sum_b x^C_{bt} \leq N^C \qquad \forall t \qquad (3)$$

In order to ensure that the hourly demand is met, the total number of fully-charged batteries should be greater than the demand at each hour. In addition, the station owner should consider the physical constraints in charging the batteries [10], [11], including the minimum charging time, minimum/maximum charging rates, state of charge limits, etc.

## 4. ILLUSTRATIVE EXAMPLE

A BSS is utilized to study the performance of the proposed model. The number of batteries owned by the BSS is considered to be 12, each with a capacity of 100 KWh. The BSS owns 4 battery chargers. Each empty battery needs to be charged for 6 hours to be fully-charged. It is assumed that there is no power limit on the BSS, i.e., 4 batteries can be charged at the same time, equivalent to the number of available chargers. The BSS schedule is studied for a 24-h horizon. Illustrative results of a possible charging schedule are provided in Table I. Each battery can have one of these four states at every hour: empty (E), charging (C), fully-charged (F), and out-of-station (O).

Batteries B6 and B7 have been charging from the previous day so they need 4 and 5 hours, respectively, to be fully-charged. On the other hand, as the station owns 4 battery chargers, some of the batteries are queued for several hours to start charging. For instance, battery B11 has remained in the empty state for 5 hours (from hour 5 to 10) until one of the chargers is opened up for this battery.

Battery B1 remains empty for the first four hours, and once one of the battery chargers becomes available, it starts charging. For the next six hours, the battery is charging until being fully-charged. This battery stays at fully-charged state because of lack of demand. As it can be seen, at hour 19, the EV owner who carries the battery B7, arrives to the station and delivers the empty battery B7 and receives battery B1. In other words, at hour 19, battery B1 is swapped with battery B7. At hour 11, fully-charged battery B7 is delivered to an EV owner, and battery B4 which was out-of-station for 9 hours, arrives to the station. Moreover, batteries B8 and B9 which have been charging from hours 1 to 6, become fully-charged at hour 7 and consequently two battery chargers are opened up for batteries B3 and B10.

This illustrative example shows how the BSS can schedule battery charging in a way that (a) constraints associated with number of chargers is closely followed, (b) there is always a fully-charged battery in the station, so as to serve customers in no time, (c) each battery is closely tracked, so the lifetime/degradation can be accurately determined, (d) it is possible that a battery stays in the station for most of the day (like B3) which is fine.

This example focuses on the feasibility of scheduling. However, an extended version of the proposed model should consider an objective to be minimized, and based on that, obtain the optimal and feasible charging schedule. The objective can be determined based on the station owner goals, such as reducing electricity payments or cooling costs, or based on the grid support objectives, such as providing reserve and flexibility to the upstream grid. In either case, the station owner has significant flexibility to schedule the available batteries while maximizing economic benefits.

Table 1. BSS scheduling.

| Batteries | Hours (1-24) | | | | | | | | | | | | | | | | | | | | | | | |
|---|---|---|---|---|---|---|---|---|---|---|---|---|---|---|---|---|---|---|---|---|---|---|---|---|
| B1  | E | E | E | E | C | C | C | C | C | C | F | F | F | F | F | F | F | F | O | O | O | O | O | O |
| B2  | E | E | E | E | E | C | C | C | C | C | C | F | F | F | F | F | F | F | F | O | O | O | O | O |
| B3  | E | E | E | E | E | E | C | C | C | C | C | C | F | F | F | F | F | F | F | F | F | O | O | O |
| B4  | F | O | O | O | O | O | O | O | O | O | E | E | C | C | C | C | C | C | F | F | F | F | F | F |
| B5  | F | F | F | F | O | O | O | O | O | O | O | O | E | C | C | C | C | C | C | F | F | F | F | F |
| B6  | C | C | C | C | F | F | O | O | O | O | O | O | O | O | E | E | C | C | C | C | C | C | F | F |
| B7  | C | C | C | C | C | F | F | F | F | F | O | O | O | O | O | O | O | O | E | C | C | C | C | C |
| B8  | C | C | C | C | C | C | F | F | F | F | F | F | O | O | O | O | O | O | O | O | E | C | C | C |
| B9  | C | C | C | C | C | C | F | F | F | F | F | F | F | O | O | O | O | O | O | O | O | O | E | C |
| B10 | O | E | E | E | E | E | C | C | C | C | C | F | F | F | F | F | F | F | F | F | F | F | F | F |
| B11 | O | O | O | O | E | E | E | E | E | C | C | C | C | C | C | F | F | F | F | F | F | F | F | F |
| B12 | O | O | O | O | O | O | E | E | E | E | C | C | C | C | C | C | F | F | F | F | F | F | F | F |

## 5. CONCLUSIONS

In this paper, the timely and viable idea of a BSS was introduced to supply power to EVs. Various involved players, such as the power system, the EV owner and the station owner, reap the benefits of the BSS. The advantages of the BSS deployment were enumerated from the perspectives of these three mentioned players. Then, from the station owner's view, a BSS scheduling model was proposed in order to charge batteries in a sequential order, while taking into account various prevailing constraints. An illustrative example was provided on a small test BSS to showcase how the proposed model would perform.

# BIBLIOGRAPHY

[1] Sh. Yang, J. Yao, T. Kang, X. Zhu, "Dynamic Operation Model of the Battery Swapping Station for EV (Electric Vehicle) in Electricity Market," (Energy, vol. 65, pp. 544–549, Feb. 2014).

[2] SW. Hadley, A. Tsvetkova, "Potential Impacts of Plug-in Hybrid Electric Vehicles on Regional Power Generation," (The Electricity Journal, vol. 22, no. 10, pp. 56–68, Dec. 2009).

[3] O. Worley, D. Klabjan, "Optimization of Battery Charging and Purchasing at Electric Vehicle Battery Swap Stations," (IEEE Vehicle Power and Propulsion Conference (VPPC), pp. 1–4, Chicago, IL, Sept. 2011).

[4] Y. Jun, H. Sun, "Battery Swap Station Location-Routing Problem with Capacitated Electric Vehicles," (Computers & Operations Research, vol. 55, pp. 217–232, march 2015).

[5] Y. Zheng, Z. Y. Dong, Y. Xu, K. Meng, J. H. Zhao, J. Qiu, "Electric Vehicle Battery Charging/Swap Stations in Distribution Systems: Comparison Study and Optimal Planning," (IEEE Transactions on Power Systems, vol. 29, no. 1, pp. 221–229, Jan. 2014).

[6] M. A. Ortega-Vazquez, F. Bouffard, V. Silva, "Electric Vehicle Aggregator/System Operator Coordination for Charging Scheduling and Services Procurement," (IEEE Transactions on Power Systems, vol. 28, no. 2, pp. 1806–1815, May 2013).

[7] Q. Dai, T. Cai, S. Duan, F. Zhao, "Stochastic Modeling and Forecasting of Load Demand for Electric Bus Battery-Swap Station," (IEEE Transactions on Power Delivery, vol. 29, no. 4, pp. 1909–1917, Aug. 2014).

[8] A. Kavousi-Fard, A. Khodaei, "Efficient Integration of Plug-in Electric Vehicles via Reconfigurable Microgrids," (Energy, vol. 111, pp.653-663, Sep.2016).

[9] S. M. Salamati, S. A. Salamati, M. Mahoor, F. R. Salmasi, "Leveraging Adaptive Model Predictive Controller for Active Cell Balancing in Li-ion Battery," (North American Power Symposium (NAPS), Morgantown, WV, Sep. 2017).

[10] A. Alanazi, A. Khodaei, "Optimal Battery Energy Storage Sizing for Reducing Wind Generation Curtailment," (IEEE PES General Meeting, Chicago, IL, July 2017).

[11] I. Alsaidan, A Khodaei, W Gao, "Determination of Optimal Size and Depth of Discharge for Battery Energy Storage in Standalone Microgrids," ( North American Power Symposium (NAPS), Denver, CO, Sep. 2016).